\documentstyle[preprint,aps]{revtex}
\begin{document}
\draft
\preprint{\vbox{
\hbox{TRI-PP-94-18}
\hbox{March 1994}
}}
\title{
Muon Number Violating Processes in Single Particle Extensions \\
of the Standard Model
}

\author{Daniel Ng and John N. Ng}
\address{
TRIUMF, 4004 Wesbrook Mall\\
Vancouver, B.C., V6T 2A3, Canada
}
\maketitle
\thispagestyle{empty}

\begin{abstract}

We study the one-loop induced muon number processes when the standard
model is minimally extended to include a $\rm SU(2)$ singlet of a
charged scalar $h^+$ and a neutral fermion $N$.  We find that $\mu
\rightarrow e \gamma$ is more sensitive for the former model whereas
$\mu-e$ conversion in nuclei for the latter.  Effects of a scalar
leptoquark $y^{1/3}$ and a heavy vector fermion $E^-$, which induce
tree level rare muon decays, are also discussed.

\end{abstract}

\newpage

Although the particle physics phenomenology  can be well
explained by the standard model (SM), most physicists believe that the
standard model is not the final theory of the nature.  There is an
indication from recent experiments at LEP suggesting that the gauge coupling
constants of the semi-simple gauge group of the standard model may meet
at high energy \cite{gut}.  This has given new impetus to the study of
grand unified theories (GUTs) and their supersymmetric variants.  Due
to the dearth of experimental information on the one hand and a
plethora of parameters in models beyond the SM such as GUT, we are
unable to discriminate among models. In general these models contain many
new particles some of which could be as light as a few hundreds
GeV.  If so, the low energy phenomenology could be affected
due to the presence of these new particles.

In this paper, we adopt a bottom up approach in unravelling physics
beyond the SM. We shall keep the SM gauge group
and study the possible of extension of the SM by adding only one new
particle to it at a time and study the low energy phenomenology of this
new particle.  By restricting ourselves to the SM gauge group, we are
led to adding either a fermion or a scalar.  Since we are adding only
one new particle, it can be a SU(2) singlet if it is
charged or a SU(2)$\times$U(1) singlet if it is neutral.
With the addition of a new particle, we construct additional renormalizable
interactions.  We do not consider non-renormalizable
interactions as they are suppressed by the high mass scale.

Let us first consider the addition of a new fermion.  At first
sight, it may appear that many possible fermions can be added.  In fact,
the choice is rather limited. For a $\rm{SU(2)\times U(1)}$ singlet, the
choice is either a right-handed neutrino $\nu_R$ or a neutral Dirac
(vectorlike) particle. For a charged $\rm{SU(2)}$ singlet,
we are limited to having either
a color singlet ($E^-$) or a color triplet ($U^{2/3}$ or $D^{-1/3}$).
Due to the anomaly consideration, these charged fermions have to be
vectorlike.  We do not consider fermions with more exotic
charges because more new scalars are required in order for them
to interact with the standard particles through the Yukawa interactions.
This would violate our philosophy of introducing only one particle.

Next we consider adding a scalar.
The choice is either a singly charged
($h^+$), fractionally charged color triplets ($y^{1/3}$) or
($x^{1/3},~x^{2/3},{\rm{or}} ~x^{4/3}$) for a $\rm{SU(2)}$ singlet,
where the particles $h^+$, $y^{1/3}$ and $x$'s
are named as the dilepton, leptoquark and diquarks because they couple
to lepton-lepton, lepton-quark and quark-quark pairs respectively.
For the case of neutral scalar, we can add a $\rm{SU(2)\times U(1)}$
singlet $h^0$.

In general, a new particle can contribute to various low energy
phenomenology.  In particular, rare muon decays is very sensitive to
such new physics because they are absolutely forbidden in the SM.
In this paper we consider how a new particle from the above list can
induce rare muon decays.  Thus, the diquarks $x$'s, and vector quarks
$U^{2/3}$ and $D^{-1/3}$ are not relevant here.
Furthermore the $h^0$, has only
interactions with the SM Higgs and gives rise to possible CP
violations in the scalar potential.  A priori it has no effect on lepton
number violating processes.  This leaves a scalar dilepton $h^+$ and
a Dirac neutrino $N$ \footnote{
If we add only one $\nu_R$, we can combine it with a linear combination
of the usual left-handed neutrinos to form a massive
Dirac neutrino and adjust the Yukawa couplings constants to give a very
small mass for this massive state.  If it has a Majorana mass, then
see-saw mass and mixing relationships are obtained.  For both cases, the
effects for rare muon decays are negligible.  When there are more than
one $\nu_R$'s, we can avoid the see-saw mass and mixing relationships
leading to significant effects on the muon number violating process
\cite{chang}. }
which could induce rare muon decays
at the one-loop level as our main focus here. The particles, $E^-$ and
$y^{1/3}$ which can induce rare decays at tree level would be
considered next.

We write the relevant effective Lagrangian relevant as
\begin{eqnarray}
{\cal L}=&&\frac{g}{2s_W m_W^2} s^2_W\lambda_1 A^\mu
   \left[ F_1\overline{e_L}(q^2\gamma_\mu-\not\hspace{-2pt}q q_\mu)\mu_L
     +F_2\overline{e_L}i\sigma_{\mu\nu}q^\nu m_\mu \mu_R \right]
      ~+~\frac{g}{c_W} \lambda_2 P_Z Z^\mu
                     \overline{e_L}~\gamma_\mu~\mu_L
\nonumber \\
  &&+\frac{g^2}{2m_W^2}\lambda_3\left[ B_e~{\overline{e_L}}~\gamma^\mu~e_L
          ~+~ B_u~{\overline{u_L}}~\gamma^\mu~u_L~+~
        B_d~{\overline{d_L}}~\gamma^\mu~d_L \right]
        {\overline{e_L}}~\gamma^\mu~\mu_L \ , \nonumber \\
&&+g \lambda_4 H \left[ C_1 {\overline{e_L}}\mu_R + C_2
{\overline{e_R}}\mu_L \right] \ ,
\label{lag}
\end{eqnarray}
where $H$ is the neutral SM Higgs scalar.
$g$ and $m_W$ are the gauge coupling constant of SU(2) and the
mass of the $W$ gauge boson. $s_W = \sin\theta_W$ where $\theta_W$ is
the Weinberg angle.  In Eq.~(\ref{lag}),
$\lambda_{1,2,3,4}$ are model dependent dimensionless parameters which
consist of products of coupling constants and mixings.
Since we do not expect direct $\mu-e-\gamma$ coupling at tree level,
$F_1$ and $F_2$ will be induced at one-loop level.  On the other hand,
tree level $\mu-e-Z$ coupling is allowed in some models and $P_Z$ would
then be unity; otherwise it will be induced at one-loop level or higher.
Similarly, for leptoquark models, $B_u$ is given as the mass-squared
ratio of $W$ to $y^{1/3}$ and $B_e=B_d=0$;
otherwise they will be given by one-loop box diagrams.
For $\mu-e-H$ vertex, we anticipate that it would be helicity suppressed,
namely, $C_1 \propto m_\mu/m_W$ and $C_2 \propto m_e/m_W$ in the models we are
interested in.
Also, the process $\mu \rightarrow 3 e$ induced
by $H$ exchange is further suppressed by the small Yukawa coupling
constant $g m_e/(2m_W)$.  However,
$\mu-e-H$ vertex will be important when there is
the flavor changing interaction involving the right-handed muon, for
instance, the supersymmetric model considered in Ref. \cite{ng}.

In this paper, we consider the muon number violating processes,
$\mu \rightarrow e \gamma$, $\mu \rightarrow 3e$ and
$\mu {\rm{Ti}} \rightarrow e {\rm{Ti}}$ which present branching
ratios are measured to be
$4.9 \times 10^{-11}$ \cite{lamf}, $1.0 \times 10^{-12}$ \cite{sindrum}
and $4.6 \times 10^{-12}$ \cite{triumf}, respectively.
These are calculated to be \cite{cheng,feinberg,shanker}
\begin{eqnarray}
\label{B1}
B(\mu \rightarrow e \gamma)&=&\frac{24\pi}{\alpha}|s^2_W\lambda_1 F_2|^2 \ ,
\\
\label{B2}
B(\mu \rightarrow 3e)&=&2L^2+R^2-4s^2_W\lambda_1F_2(2L+R)+
  4(s^2_W\lambda_1F_2)^2\left(4\ln\frac{m_\mu}{2m_e}-\frac{13}{6}\right) \ , \\
\label{B3}
B(\mu-e)&=&\frac{1}{\Gamma_0} \frac{\alpha^3 G_F^2 m_\mu^5}
       {\pi^2} \frac{Z_{eff}^4}{Z} |F(-m_\mu^2)|^2 |Q_W|^2 \ ,
\end{eqnarray}
with
\begin{eqnarray}
\label{L}
L =&& s^2_W\lambda_1 F_1+\left(-1+2s_W^2 \right)\lambda_2 P_Z -\lambda_3 B_e \
, \\
\label{R}
R =&& s^2_W\lambda_1 F_1+ 2s_W^2\lambda_2 P_Z \ , \\
\label{QW}
Q_W =&&\left[\frac{2}{3}s^2_W\lambda_1(F_2-F_1)+(\frac{1}{2}-
   \frac{4}{3}s^2_W)\lambda_2P_Z -\frac{1}{2}\lambda_3B_u
\right] (2Z+N) \nonumber \\
 &&+\left[-\frac{1}{3}s^2_W\lambda_1(F_2-F_1)+(-\frac{1}{2}+
   \frac{2}{3}s^2_W)\lambda_2P_Z -\frac{1}{2}\lambda_3B_d
\right] (Z+2N) \ ,
\end{eqnarray}
where $G_F$ is the Fermi four-fermion coupling constant.
$F(-m_\mu^2)= 0.54$\cite{dreher}  and $Z_{eff}=17.6$ \cite{ford}
are the nuclear form factor and the
effective atomic number for the nuclei $\rm{^{48}_{22} Ti}$.
$\Gamma_0 = (2.590\pm 0.12)\times 10^6~sec^{-1}$ \cite{suzuki}
is the muon capture rate for Ti.

{\underline{\it A model with a charged scalar dilepton $h^+$}}.
When a new charged scalar $h^+$
is introduced to the SM, new Yukawa interactions, which are given by
\begin{equation}
\label{h_yukawa}
-{\cal L}_{\rm{Y}}({\rm{new}})~=~f_1 ( \nu_e \mu - e \nu_\mu) h^+~+~
   f_2 ( \nu_e \tau - e \nu_\tau) h^+~+~
    f_3 ( \nu_\mu \tau - \mu \nu_\tau) h^+~+~H.c. \ ,
\end{equation}
are allowed.  A new scalar potential depicting the interaction of $h^+$
with the SM Higgs doublet field $\Phi$ can be constructed and is given by
\begin{equation}
V~=~m^2 |h^+|^2 + \lambda |h^+|^4 + a |h^+|^2 \Phi^\dagger \Phi +
V(\Phi) \ ,
\end{equation}
We can see from Eq.~(\ref{h_yukawa}) that adding a
$h^+$ will break the family lepton numbers, ($L_e$, $L_\mu$ and
$L_\tau$) but preserve the total lepton number, $L$.  In addition, $h^+$
which is named as dilepton
carries $L = -2$.  This model is a simplified version of the Zee model
\cite{zee} in which two Higgs doublets are introduced.   The main
motivation there is to generation neutrino masses.

Owing to Eq.~(\ref{h_yukawa}), the $e$--$\mu$--$\tau$ universality is
broken if $f_1 \not \hspace{-1pt}= f_2 \not \hspace{-1pt} =f_3$.
In particluar, Fermi coupling constant $G_F$, which is extracted from
the muon lifetime, will be modified to be
\begin{equation}
\label{fermi}
\frac{G_F}{\sqrt{2}}~=~\frac{g^2}{8m_W^2}
  \left[ 1+ \frac{f_1^2}{g^2 y} \right] + O(\frac{f_{2,3}^4}{g^4 y^2})
\end{equation}
due to the exchange of $h^+$, where $y=m_h^2 /m_W^2$.
The best constraint on $f_1$ can be obtained from the nuclear
beta decay and $K_{e3}$ decay.
Normalized to the muon decay, the CKM elements
become $(|V_{ud}|^2+|V_{us}|^2+|V_{ub}|^2)
\left[1-2{f_1^2}/{g^2 y}\right] = 0.9979\pm0.0021$ \cite{databook}.
By the unitarity of the CKM matrix, we obtain ${f_1^2}/y \leq
2.1\times10^{-3}$ at $90\%$ C.L..
In addition, the $f_2$ and $f_3$ terms will break the $e-\mu$
universality in the $\tau$ decays $\tau \rightarrow e \nu \nu$ and
$\tau \rightarrow \mu \nu \nu$.
Experimentally, the updated world average for these two decays modes are
\cite{weinstein} $17.77\pm0.15\%$ and $17.48\pm0.18\%$ respectively.
Combining the constraints on $f_1$ and $f_2$ obtained from $\mu-\tau$
universality violation in the decays $\mu\rightarrow e\nu\nu$ and
$\tau\rightarrow e\nu\nu$ \cite{marciano} with the updated world averages
of the tau lepton mass
and lifetime, $m_\tau=1770.0\pm0.4~\rm MeV$ and
$\tau_\tau=(295.9\pm3.3)\times10^{-15}\rm~s$ respectively, we obtain
that the ratios $f_{1,2}^2/y~(f_3^2/y)$ are bounded to be on the order
$10^{-3}~(10^{-2})$ or less.

Since the family lepton numbers are explicitly violated by $h^+$,
rare muon decays, such as $\mu \rightarrow e\gamma$, $\mu \rightarrow 3
e$ as well as $\mu^--e^-$ conversion in nuclei, are allowed by one-loop
quantum corrections due to $h^+$.  Let us first consider the usual
photon penguin diagrams of $\mu-e$ transition, see
Fig.~\ref{fig:penguin}.
Here we have $\lambda_1~=~(f_2f_3)/(8\pi^2)$ and
the corresponding charge radius and magnetic moment terms, $F_1$ and
$F_2$, are given by
\begin{equation}
\label{F1F2h}
F_1~=~-\frac{1}{18} \frac{1}{y} ,
F_2~=~-\frac{1}{12} \frac{1}{y} \ .
\end{equation}
respectively.

Since $m_\mu \ll m_Z$, it is a good approximation to neglect the
external momenta for the processes we are considering.
The $Z$ penguin diagrams are then explicitly given as
\begin{eqnarray}
\label{Za}
P_Z(a)~&=&~\sin^2\theta_W \left[
  ~\frac{1}{\epsilon}~+~\frac{1}{4}~-~\frac{1}{2}\ln{m_h^2} \right] \ , \\
\label{Zb}
P_Z(b)~&=&~ \left[
  -~\frac{1}{2\epsilon}~-~\frac{1}{8}+\frac{1}{4}\ln{m_h^2}\right] \ , \\
\label{Zc}
P_Z(c+d)~&=&~(-\frac{1}{2}+\sin^2\theta_W)
 \left[ -\frac{1}{\epsilon}~-~\frac{1}{4}~+~\frac{1}{2}\ln{m_h^2}\right] \ .
\end{eqnarray}
Summing Eq.~(\ref{Za}) to (\ref{Zc}),  we
find that the effective $\mu-e-Z$ vertex vanishes, in the approximation
that neglects the external momenta.
The corrections due to external momenta
for the processes such as $\mu \rightarrow 3 e$ and $\mu-e$
conversion are suppressed by a small factor $m_\mu^2/m_Z^2$.
Hence, $Z$ penguin diagrams can be neglected in this model.

For the two $h^+$ exchange box diagrams, we obtain
$\lambda_3=\lambda_1(f_1^2+f_2^2)/g^2$, $B_e = -1/(4y)$ and $B_u=B_d=0$.
However, the contributions are small unless the coupling $f$'s are on
the order of $g$.

Let us now consider the process $\mu \rightarrow e \gamma$ which the
experimental bound on the branching ratio is $4.9\times10^{-11}$
\cite{lamf}.  This translates into a better constraint for $f_{2,3}$,
namely
\begin{equation}
\frac{f_2f_3}{y} < 2.8\times10^{-4} \ .
\end{equation}

To study the relative importance of the different rare muon decays, we
construct the branching ratios for $\mu \rightarrow 3 e$ and $\mu-e$
conversion in Ti with respect to $\mu \rightarrow e \gamma$.
Explicitly, we obtain
\begin{eqnarray}
\label{R1}
R_1~&&=~\frac{B(\mu \rightarrow 3 e)}{B(\mu \rightarrow e \gamma)}~=~
\frac{\alpha}{24\pi}\frac{3F_1^2-12F_1F_2+
  4F_2^2(4\ln\frac{m_\mu}{2m_e}-\frac{13}{6})}{F_2^2} ~=~5.7\times10^{-3}\ ,
\\
\label{R2}
R_2~&&=~\frac{B(\mu {\rm{Ti}} \rightarrow e {\rm{Ti}})}
     {B(\mu \rightarrow e \gamma)}
      ~=~5.3\times10^{-6} \frac{|(F_2-F_1)(Z+\frac{1}{3}N)|^2}{|F_2|^2}
    ~=~ 2.7\times10^{-4}\ .
\end{eqnarray}
Clearly, $\mu \rightarrow e \gamma$ is the best probe of the charged
scalar singlet model.

{\underline{\it A model with a neutral Dirac fermion N}}.
In many GUT models, there exist heavy neutral fermions.  For example,
the $\rm{E_6}$ GUT model which was first considered in Ref. \cite{ramond}
has fermion states for each generations placed in a $\underline{27}$
representation.  Thus, there are additional
$12$ fermion states in addition to the
$15$ SM fermion states.  The new particles, given in
terms of left-handed chirality,  are
color singlet fermions ($E$, $E^c$, $\nu_E$, $N_E^c$, $N^c$, $n$) and
color triplet fermions ($D$, $D^c$).   The representation of the new
particles under the standard $\rm{SU(2)\times U(1)}$ depends on the
$\rm{E_6}$ symmetry breaking scheme.  Motivated by this GUT, we first
consider the neutral particle $N$.

$N$, which is a $\rm{SU(2)\times U(1)}$ singlet neutral fermion, can be
either a Dirac or Majorana particle.  The phenomenology of having a
Majorana particle has been extensively studied in Ref. \cite{chang}.  Here, we
consider $N$ being a Dirac particle.  Since it is a $\rm{SU(2)\times
U(1)}$ singlet particle, a gauge invariant mass term
\begin{equation}
-{\cal L}_M~=~M_N \overline{N}_L N_R~+~H.c. \ ,
\end{equation}
is allowed.  The new Yukawa interactions are given as
\begin{equation}
\label{Nyukawa}
-{\cal L}_Y({\rm{new}})~=~\sum_{\alpha=e,\mu,\tau} f_\alpha
\pmatrix{{\overline{\nu}_\alpha}_L&\overline{\alpha}_L}N_R
\pmatrix{\overline{\phi^0}\cr-\phi^-}~+~H.c. \ ,
\end{equation}
where the charged lepton states are defined to be the mass eigenstates.
Owing to the fact that the usual neutrinos couple to the massive
neutrino shown in Eq.~(\ref{Nyukawa}), the definition of the massless
neutrinos is not arbitrary.  The flavor states are related to the mass
eigenstates by ${(\nu_e,\nu_\mu,\nu_\tau,N)_L}^T$ = $ {\cal O}~
{(\nu_1,\nu_2,\nu_3,\nu_4)_L}^T$ where ${\cal O}$ is given by
\begin{equation}
\label{O}
{\cal O}~=~\pmatrix{c_1&s_1c_2&s_1s_2c_3&s_1s_2s_3\cr
     -s_1&c_1c_2&c_1s_2c_3&c_1s_2s_3 \cr 0&-s_2&c_2c_3&c_2s_3 \cr
       0&0&-s_3&c_3} \ .
\end{equation}
$s_i$ are abbreviation of $\sin\theta_i$ which are given as
$s_1=f_e/\sqrt{f_e^2+f_\mu^2}$, $s_2=\sqrt{f_e^2+f_\mu^2}/f$
and $s_3= m_D/M_N$ where $f = \sqrt{f_e^2+f_\mu^2+f_\tau^2}$ and
$m_D = f <\phi^0>$.
$\nu_1$, $\nu_2$ and $\nu_3$ remain massless;
whereas ${\nu_4}_L$ combines with $N_R$ to form a Dirac neutrino
with a mass equal to $\sqrt{m_D^2+M_N^2}$.  Therefore, the existence
of the $N$ will break the individual lepton flavor number conservation,
leading to rare muon decays such as
$\mu \rightarrow e \gamma$, $\mu \rightarrow 3 e$ and $\mu-e$
conversion in nuclei.   However, the total lepton number remains conserved.

The presence of $N$ would lead to the flavor changing neutrino-$Z$ gauge
coupling \cite{escobar} which is given by
\begin{eqnarray}
\label{Znn}
{\cal L}_{Z\bar \nu \nu}~=&&~\frac{g}{4c_W} Z_\mu \left[
{\overline{\nu_1}_L}\gamma^\mu{\nu_1}_L
+{\overline{\nu_2}_L}\gamma^\mu{\nu_2}_L
+c_3^2{\overline{\nu_3}_L}\gamma^\mu{\nu_3}_L \right. \nonumber \\
&&+\left. s_3^2{\overline{\nu_4}_L}\gamma^\mu{\nu_4}_L
+s_3c_3{\overline{\nu_3}_L}\gamma^\mu{\nu_4}_L
+s_3c_3{\overline{\nu_4}_L}\gamma^\mu{\nu_3}_L \right] \ .
\end{eqnarray}
Since we expect the new particle to come from the higher mass
scale, it is reasonable to assume $m_4 > m_Z$.  Thus, the invisible width of
$Z$ would be reduced due to the smaller coupling to $\nu_3$.
The number of light neutrino species in the $Z$ decay is given by
\begin{equation}
\label{Nn}
N_\nu=2+(1-s_3^2)^2 \ .
\end{equation}
$N_\nu$ is experimentally measured to be $2.980\pm0.027$ at the LEP
\cite{schaile}, leading to
\begin{equation}
\label{s3bound}
{s_3}^2 \leq 3.05 \times10^{-2} (90 \% {\rm{C.L.}}) \ .
\end{equation}

The neutrino mixings would also contribute to the $e-\mu-\tau$
universality violation \cite{pilaftsis,chang}, similar to our previous
discussion.  However, the constraint
Eq.~(\ref{s3bound}) is more appropriate for
our discussions of rare muon decays.

Although the neutrino flavor is violated at tree level in the $Z$ gauge
interaction given by Eq.~(\ref{Znn}), rare muon decays are induced at
one-loop level.  From Eq.~(\ref{O}) and consideration of the $\gamma$
penguin, we extract $\lambda_1=g^2/(16\pi^2){\cal O}_{\mu 4}^\ast
{\cal O}_{e 4}$.  The charge radius and the magnetic moment terms
for the photon $\mu-e$ transition \cite{inami} are
\begin{eqnarray}
\label{F1N}
F_1&=&\frac{x_4(12+x_4-7x_4^2)}{12(x_4-1)^3}+
	\frac{x_4^2(-12+10x_4-x_4^2)}{6(x_4-1)^4}\ln{x_4} \ , \\
\label{F2N}
F_2&=&\frac{x_4(1-5x_4-2x_4^2)}{4(x_4-1)^3}+
	\frac{3x_4^3}{2(x_4-1)^4}\ln{x_4}  \ ,
\end{eqnarray}
where $x_4 = m_4^2/m_W^2$.
For the $Z$ penguin diagram, care has to be taken due to the flavor
changing coupling, and the final result can be simply written as
\begin{equation}
\label{PZN}
P_Z~=~\left[-\frac{5x_4}{2(x_4-1)}+
   \frac{3x_4^2+2x_4}{2(x_4-1)^2}\ln{x_4}\right]
     ~+~\frac{1}{2}s_3^2 x_4 \ ,
\end{equation}
and $\lambda_2=g^2/(32\pi^2){\cal O}_{\mu 4}^\ast {\cal O}_{e 4}$.
The integrals $B_e$, $B_u$ and $B_d$ of the usual $W$--boson exchange box
diagrams for $\mu\rightarrow 3 e$ and $\mu-e$ conversion
are given by
\begin{eqnarray}
\label{BeN}
B_e~&=&~\left[\frac{x_4}{x_4-1}-\frac{x_4}{(x_4-1)^2}\ln{x_4}\right]
    ~+~|{\cal O}_{e4}|^2 \left[\frac{-4x_4+11x_4^2-x_4^3}{4(x_4-1)^2}-
     \frac{3x_4^3}{2(x_4-1)^3}\ln{x_4}\right] \, \\
B_u~&=&~\frac{4x_4}{x_4-1}-\frac{4x_4}{(x_4-1)^2}\ln{x_4} \, \\
B_d~&=&~\frac{x_4}{x_4-1}-\frac{x_4}{(x_4-1)^2}\ln{x_4} \,
\end{eqnarray}
and $\lambda_3 = \lambda_2$.

Let us first apply the above expressions to
$\mu\rightarrow e\gamma$.  Using the experimental
bound given in Ref.~\cite{lamf}, we obtain
\begin{equation}
|\lambda_1 F_2 | \leq 3.0\times10^{-7}.
\end{equation}
This translates into $\lambda_1 \leq
2.4\times10^{-6}~(6.7\times10^{-7})$ for $m_4 = m_W~(10~m_W)$.

Next we consider the process $\mu\rightarrow 3e$.
{}From the expressions for the branching ratios given in Eqs. (\ref{B1})
and (\ref{B2}), the ratio $R_1~=~{B(\mu\rightarrow 3e)}/
{B(\mu\rightarrow e\gamma)}$ is proportional to a
small factor $\alpha/(24\pi)=1\times10^{-4}$.  Naively, one would expect
the branching ratio for $\mu\rightarrow 3e$ to be always very small.
In fact, this is not the case because the process $\mu\rightarrow 3e$
receives large contributions from the $Z$ penguin diagrams.
When we include also the box diagram contributions, we obtain
$R_1 \leq 0.03~(0.10)$ \cite{comment} for $m_4~=~m_W~(10~m_W)$.
Since the present experiment sensitivity for the experiment
$\mu\rightarrow 3e$ is $50$
times better than that of $\mu\rightarrow e\gamma$, the former
experiment is better in probing the physics of
a heavy Dirac neutral fermion $N$.

For the $\mu{\rm{Ti}}\rightarrow e{\rm{Ti}}$ experiment, the ratio of
the branching ratio relative to that of $\mu\rightarrow e\gamma$
can be written as
\begin{equation}
R_2~=~5.1\times10^{-5} \frac{|Q_W|^2}{|s^2_W\lambda_1F_2|^2} \ .
\end{equation}
Putting in limits for the various parameters,
we obtain $R_2 \leq 12~(6.3)$ \cite{comment} for
$m_4~=~m_W~(10~m_W)$.  Therefore, this experiment is far better than the
other two.

{\underline{\it Tree Level Rare Muon decays}} As we have discussed
previously, a new heavy lepton $E^-$ can exist in GUT models
such as $\rm{E_6}$.  Due to anomaly considerations,
$E$ has to be vectorlike under the SM gauge group.
Thus the Yukawa interactions in the lepton sector and the gauge
invariant mass term for $E$ are given as
\begin{equation}
\label{YE}
-{\cal L}~=~h_{ij}\overline{L^i}e_R^j\Phi~+~
    f_{i4}\overline{L^i}E_R\Phi~+~M_E \overline{E_L}E_R~+~H.c. \ ,
\end{equation}
where $i,j = 1,2,3$ are the family indices.  $L ~(e_R)$ is the usual SU(2)
doublet (singlet) leptons.  When the neutral Higgs acquires vacuum
expectation values $<\phi^0>$,  the charged leptons get a mass matrix
$M_l= (h+f)<\phi^0> + M_E~diag(0,0,0,1)$.   We can diagonalize this mass
matrix by a bi-unitary transformation $U_L^\dagger M_l U_R $.  As shown
in Eq.~(\ref{YE}), the charged lepton masses come from two sources,
namely $<\phi^0>$ and $M_E$.  Thus, the tree-level
lepton flavor changing Higgs vertex
is induced.  However, it is suppressed by $m_\mu/m_W$.

On the other hand, flavor changing $Z$ coupling is also induced by $E$
because it has different gauge transformation in the left-handed sector.
This leads to large tree level contributions to $\mu
\rightarrow 3e$ and $\mu-e$ conversion.  At the tree level we
have $\lambda_1=\lambda_3=0$, and $\lambda_2~=~(1/2){U_L}^\ast_{E e}
{U_L}_{E\mu}$ with $P_Z~=~1$.
In this case, both processes $\mu{\rm{Ti}}\rightarrow e{\rm{Ti}}$ and
$\mu \rightarrow 3 e$ proceed through the lepton flavor changing $Z$
coupling at the tree level.  Thus, the ratio
${B(\mu{\rm{Ti}}\rightarrow e{\rm{Ti}})}/{B(\mu \rightarrow 3 e)}$
is approximately equal to $10$.  This implies that
that the best probe for the tree level flavor changing $Z$ coupling
induced by $E^-$ is the $\mu-e$ conversion experiment and we arrive
at the bound ${U_L}^\ast_{Ee} {U_L}_{E\mu} < 1.5 \times 10^{-6}$

The second example for this class of model is the existence of a leptoquark
$y^{1/3}$ which induces a new Yukawa interaction $f_{ij} L^i Q^j y^{1/3}
+ H.c. $, where $L$ and $Q$ are the lepton and quark doublets where
$i$ and $j$ are the generation indices.  In this case,
$\lambda_1=\lambda_2=B_e=B_d=0$, and $\lambda_3~=~f_{\mu u}f^\ast_{e u}/g^2$
and $B_u=-(m_W/m)^2$, where $m$ is the mass for the $y^{1/3}$.
Therefore, in this model only $\mu-e$ conversion is allowed at the
tree level by the exchange of $y^{1/3}$, leading to a stringent constraint
$f_{\mu u}f^\ast_{e u} (m_W/m)^2 < 1.1 \times 10^{-7}$.

In conclusion, we have studied the effects on the muon number violating
decays induced by adding one new particle to the SM.
When the rare muon decays such as $\mu \rightarrow e
\gamma$, $\mu \rightarrow 3 e$ and $\mu{\rm{Ti}}\rightarrow e{\rm{Ti}}$
are induced by new physics at the one-loop level,
one would expect that the first process to be dominant because it
is kinematically more favorable.  However, this is not necessary always true.
We find that for the case of a charged scalar $h^+$,
$\mu \rightarrow e \gamma$ is the most sensitive experiment
to probe a charged scalar singlet $h^+$.  On the other hand,
$\mu-e$ conversion in nuclei is the best probe for a heavy
Dirac neutrino $N$ because of the large contribution coming from the $Z$
penguin diagrams.  Tree level effects
are also possible by adding an charged vectorlike lepton $E^-$ or
a leptoquark scalar $y^{1/3}$.  Again, the most sensitive experiment is
$\mu-e$ conversion in nuclei.  In table \ref{table},
we present the constraints for $\lambda$'s obtained from
these three muon number violating processes with the masses for
the new particles taken to be $m_W$ and $10~m_W$.
For completeness, we include the result
given in Ref.~\cite{chang} for the case of adding right-handed Majorana
neutrinos.

\vspace{1cm}
This work was supported in part by the Natural Science and Engineering
Council of Canada.

\newpage
\vspace{1cm}
\begin{table}
\caption{The constraints for $\lambda$'s obtained from the processes
$\mu \rightarrow e~\gamma$, $\mu\rightarrow 3 e$,
and $\mu~{\rm{Ti}} \rightarrow e~{\rm{Ti}}$, where the masses for
the new particles are taken to be from $m_W~(10~m_W)$.}
\vspace{1cm}
\begin{tabular}{|c|c|c|c|} \hline
Model
& $B(\mu \rightarrow e \gamma)$
& $B(\mu \rightarrow 3 e)$
& $B(\mu {\rm{Ti}}\rightarrow e {\rm{Ti}})$
\\
\hline
$h^+$
& $\lambda_1 \stackrel{<}{\sim} 3.6\times10^{-6}(3.6\times10^{-4})$
& $\lambda_1 \stackrel{<}{\sim} 6.7\times10^{-6}(6.7\times10^{-4})$
& $\lambda_1 \stackrel{<}{\sim} 6.8\times10^{-5}(6.8\times10^{-3})$ \\
\hline
$N$
& $\lambda_1 \stackrel{<}{\sim} 2.4\times10^{-6}(6.7\times10^{-7})$
& $\lambda_1 \stackrel{<}{\sim} 1.8\times10^{-6}(2.8\times10^{-7})$
& $\lambda_1 \stackrel{<}{\sim} 2.1\times10^{-7}(8.0\times10^{-8})$ \\
\hline
$E^-$
& n/a
& $\lambda_2\stackrel{<}{\sim} 1.1\times10^{-6}$
& $\lambda_2\stackrel{<}{\sim} 7.7\times10^{-7}$  \\
\hline
$y^{1/3}$
& n/a
& n/a
& $\lambda_3\stackrel{<}{\sim} 2.7\times10^{-7}(2.7\times10^{-5})$\\
\hline
$\nu_R~^{\rm a}$
& $\lambda_1 \stackrel{<}{\sim} 2.4\times10^{-6}(6.7\times10^{-7})$
& $\lambda_1 \stackrel{<}{\sim} 1.9\times10^{-6}(1.5\times10^{-7})$
& $\lambda_1 \stackrel{<}{\sim} 2.1\times10^{-7}(6.9\times10^{-8})$ \\
\end{tabular}
\label{table}
\tablenotetext[1]{
we include the result for the Majorana neutrino model studied in
Ref.~\cite{chang}}
\end{table}

\begin{figure}
\caption{$\gamma$ and $Z$ penguin diagrams for $\mu$--$e$ transition for
a scalar $h^+$ model}
\label{fig:penguin}
\end{figure}


\begin{references}
\bibitem{gut} U. Amaldi, W. de Boer and H. Furstenan, Phys. Lett. B{\bf
260}, 447 (1991).
\bibitem{ng} D. Ng and J.N. Ng, Phys. Lett. B{\bf 320}, 181 (1994).
\bibitem{lamf} R.D. Bolton, et al., LAMF, Phys. Rev. D{\bf 38} 2077
(1988).
\bibitem{sindrum} U. Bellgardt, et al., Nucl. Phys. B{\bf 299}, 1
(1988).
\bibitem{triumf} S. Ahmad, et al., TRIUMF, Phys. Rev. D{\bf 38}, 2102
(1988).
\bibitem{cheng} T.P. Cheng and L.F. Li, Phys. Rev. D{\bf 16} 1425
(1977), ibid D{\bf 44}, 1502 (1991).
\bibitem{feinberg} G. Feinberg and S. Weinberg, Phys. Rev. Lett. {\bf
3}, 111, 244(E) (1959).
\bibitem{shanker} O. Shanker, Phys. Rev. D{\bf 20}, 1608 (1979).
\bibitem{dreher} B. Dreher, et. al., Nucl. Phys. A{\bf 235}, 219 (1974);
B. Frois and C.N. Papanicolas, Ann. Rev. Nucl. Sci. {\bf 37}, 133
(1987).
\bibitem{ford} K.W. Ford and J.G. Wills, Nucl. Phys. {\bf 35}, 295
(1962); R. Pla and J. Bernabeu, An. Fis. {\bf 67}, 455 (1971).
\bibitem{suzuki} T. Suzuki, D.F. Measday and J.P. Roalsvig, Phys. Rev.
C{\bf 35}, 2212 (1987).
\bibitem{zee} A. Zee, Phys. Lett. B{\bf 93}, 289 (1980).
\bibitem{sirlin} G. Degrassi, S. Fanchiotti, and A. Sirlin, Nucl. Phys.
\bibitem{schaile} D. Schaile, Report No. CERN-PPE/93-213.
\bibitem{databook} Particle Data Group, Phys. Rev. D {\bf 45}, part II,
June (1992).
\bibitem{weinstein} A.L. Weinstein, and R. Stroynowski, CALT-68-1853,
February 1993.
\bibitem{marciano} W.J. Marciano and A. Sirlin, Phys. Rev. Lett. {\bf
61}, 1815 (1988).
\bibitem{escobar} C. Escobar, O. Peres, V. Pleitez and R. Funchal, Phys.
Rev. D {\bf R1747}, 1993.
\bibitem{pilaftsis} J. Bernabeu, J. Korner, A. Pilaftsis and K.
Schilcher, Phys. Rev. Lett. {\bf 71}, 2695 (1993).
\bibitem{chang} L.N. Chang, D. Ng and J.N. Ng, {\it Phenomenological
Consequences of singlet Neutrinos}, TRI-PP-94-1, VPI-IHEP-94-1, February
1994.
\bibitem{ramond} F. Gursey, P. Ramond and P. Sikivie, Phys. Lett. B{\bf
60}, 177 (1976).
\bibitem{inami} T. Inami and C.S. Lim, Prog. Theor. Phys. {\bf 65}, 297
(1981); ibid. {\bf 65}, 1772 (1981).
\bibitem{comment} These are the maximum values allowed with the
constraints $s^2_3 \le 3.05\times10^{-2}$ obtained in Eq.~(\ref{s3bound}).

\end{references}
\end{document}